\documentclass[runningheads]{llncs}
\usepackage{makeidx,multirow,tikz}

\begin{document}

\frontmatter

\pagestyle{headings}

\mainmatter

\title{Towards Optimal Sorting of 16 Elements}
\titlerunning{Towards Optimal Sorting of 16 Elements}

\author{Marcin Peczarski}
\authorrunning{Marcin Peczarski}

\tocauthor{Marcin Peczarski}
\institute{Institute of Informatics, University of Warsaw,\\
ul.\ Banacha 2, 02-097 Warszawa, Poland,\\
\email{marpe@mimuw.edu.pl}}

\maketitle

\begin{abstract}
One of the fundamental problem in the theory of sorting is to find the
pessimistic number of comparisons sufficient to sort a given number of elements.
Currently 16 is the lowest number of elements for which we do not know the exact
value. We know that 46 comparisons suffices and that 44 do not. There is an open
question if 45 comparisons are sufficient. We present an attempt to resolve that
problem by performing an exhaustive computer search. We also present an
algorithm for counting linear extensions which substantially speeds up
computations.
\end{abstract}

\section{Introduction}

We consider sorting by comparisons. One of the fundamental problem in that area
is to find the pessimistic number $S(n)$ of comparisons sufficient to sort $n$
elements. Steinhaus posed this problem in~\cite{Steinhaus}. Knuth considered it
in~\cite{Knuth}. From the \emph{information-theoretic lower bound}, further
denoted by ITLB, we know that $S(n)\ge\lceil\log_2 n!\rceil=C(n)$. Ford and
Johnson  discovered \cite{FordJohnson} an algorithm, further denoted by FJA,
which nearly and sometimes even exactly matches $C(n)$. Let $F(n)$ be the
pessimistic number of comparisons in the FJA. It holds $S(n)=F(n)=C(n)$ for $n
\le 11$ and $n=20,21$. The FJA does not achieve the ITLB for $12 \le n \le 19$
and infinitely many $n\ge22$. Carrying an exhaustive computer search, Wells
discovered in 1965 \cite{Wells65,Wells71} that the FJA is optimal for 12
elements and $S(12)=F(12)=C(12)+1=30$. Kasai et al.\ \cite{Kasai} computed
$S(13)=F(13)=C(13)+1=34$ in 1994, but that result was not widely known. It was
discovered again a few years later \cite{Peczarski02}, independently, extending
the Wells method. Further improvement of the method led to show in years
2003--2004 \cite{Peczarski04,Peczarski07} that it holds $S(n)=F(n)=C(n)+1$ for
$n=14,15,22$, similarly.

In this paper we consider the case $n=16$. This is now the lowest number of
elements for which we do not know the exact value of $S(n)$. The previous
results could suggest that $S(16)=F(16)=C(16)+1=46$. However Knuth conjectures
that $S(16)=C(16)=45$. He does not believe that the FJA is optimal for 16
elements. He wrote \cite{Knuth}: ``There must be a way to improve upon this!''
We present recently obtained results\footnote{The results presented in this
paper are obtained using computer resources of the Interdisciplinary Centre for
Mathematical and Computational Modelling (ICM), University of Warsaw.} aiming
to compute the value of $S(16)$. It is very unlikely that someone will find it
by pure theoretical consideration. It seems that the only promising way leads
by performing an exhaustive computer search supported by cleaver heuristic.

The paper is organized as follows. In Sect.\ \ref{sec:notation} we introduce
notation used throughout the paper. In Sect.\ \ref{sec:algorithm} we briefly
describe the algorithm we use to resolve if there exists a sorting algorithm for
a given number of elements and comparisons. We analyse why the ITLB is not
achieved for 13, 14 and 15 elements in Sect.\ \ref{sec:prevcases}. We present
the newest results for 16 elements in Sect.\ \ref{sec:attempt}. In Sect.\
\ref{sec:complexity} we compare the computation complexity of the previous cases
and the case of 16 elements. Finally, in Sect.\ \ref{sec:counting}, we present
the algorithm for counting linear extensions which substantially improves the
algorithm from Sect.\ \ref{sec:algorithm}.

\section{Notation}\label{sec:notation}

We denote by $U=\{u_0,u_1,\dots,u_{n-1}\}$ an $n$-element set to be sorted.
Sorting of the set $U$ is represented as a sequence of posets
$(P_c=(U,R_c))_{c=0,1,\dots,C}$, where $R_c$ is a partial order relation over a
set $U$. Sorting starts from the total disorder $P_0=(U,R_0)$, where
$R_0=\{(u,u):u\in U\}$. After performing $c$ comparisons we obtain a poset
$P_c=(U,R_c)$. Sorting should end with a linear order $P_C$. Assume that
elements $u_j$ and $u_k$ are being compared in step $c$. Without loss of
generality we can assume that $(u_j,u_k)\not\in R_{c-1}$ and $(u_k,u_j)\not\in
R_{c-1}$. Suppose the answer to the comparison is that \emph{element $u_j$ is
less than element $u_k$}. Then we obtain the next poset $P_c=(U,R_c)$, where
the relation $R_c$ is the transitive closure of the relation
$R_{c-1}\cup\{(u_j,u_k)\}$. We denote this by $P_c=P_{c-1}+u_j u_k$.

By $e(P)$ we denote the number of linear extensions of a poset $P=(U,R)$. We
assume that $e(P+u_j u_k)=e(P)$ and $e(P+u_k u_j)=0$ if elements $u_j$, $u_k$
are in relation, i.e., if $(u_j,u_k)\in R$.

\section{The Algorithm}\label{sec:algorithm}

In this section we remember briefly the algorithm which answers if sorting of a
given poset $P_0$ can be finished in $C$ comparisons. The algorithm was
invented in \cite{Wells65,Wells71} and improved in \cite{Peczarski02} and later
in \cite{Peczarski04}. We present the next improvement to the algorithm in
Sect.\ \ref{sec:counting}. The algorithm has two phases: forward steps and
backward steps.

In the forward steps we consider a sequence of sets $({\cal
S}_c)_{c=0,1,\dots,C}$. The set ${\cal S}_0$ contains only the poset $P_0$. In
step $c$ we construct the set ${\cal S}_c$ from the set ${\cal S}_{c-1}$. Every
poset $P\in{\cal S}_{c-1}$ is examined for every unrelated pair $(u_j,u_k)$ in
order to verify whether it can be sorted in the remaining $C-c+1$ comparisons.
As the result of the comparison of $u_j$ and $u_k$ one can get one of two
posets $P_1=P+u_j u_k$ or $P_2=P+u_k u_j$. If the number of linear extensions
of $P_1$ or $P_2$ exceeds $2^{C-c}$ then by the ITLB it cannot be sorted in the
remaining $C-c$ comparisons. It follows that in this case, in order to finish
sorting in $C-c+1$ comparisons, elements $u_j$ and $u_k$ should not be compared
in step $c$. If the number of linear extensions of both $P_1$ and $P_2$ do not
exceed $2^{C-c}$ then we store one of them in the set ${\cal S}_c$, namely that
with greater number of linear extensions. If both have the same number of
linear extensions we choose $P_1$ arbitrarily. We do not store isomorphic
posets or a poset which dual poset is isomorphic to some already stored poset.

If some set ${\cal S}_c$ in the sequence appears to be empty then we conclude
that the poset $P_0$ cannot be sorted in $C$ comparisons. Such results are
received for 12 and 22 elements and $C=C(n)$ \cite{Peczarski04}, where the set
${\cal S}_{23}$ and ${\cal S}_{40}$ is empty, respectively. Wells reported
\cite{Wells71} that for $n=12$ only the set ${\cal S}_{24}$ is empty. Those
results mean that $S(n) > C(n)$ for $n=12,22$. If the set ${\cal S}_C$ is not
empty after performing forwards steps, we cannot conclude about sorting of the
poset $P_0$. In that case we continue with backward steps.

In the backward steps we consider the sequence of sets $({\cal
S}^*_c)_{c=0,1,\dots,C}$. We start with the set ${\cal S}^*_C={\cal S}_C$ which
contains only a linear order of the set $U$. In step $c$, where
$c=C-1,C-2,\dots,0$, we construct the set ${\cal S}^*_c$ from the set ${\cal
S}^*_{c+1}$. The set ${\cal S}^*_c$ is a subset of the set ${\cal S}_c$ and
contains only posets which can be sorted in the remaining $C-c$ comparisons.
Poset $P\in{\cal S}_c$ is stored in ${\cal S}^*_c$ iff there exists in $P$ a
pair of unrelated elements $(u_j,u_k)$ such that poset $P_1=P+u_j u_k$ or poset
$P_2=P+u_k u_j$ belongs to the set ${\cal S}^*_{c+1}$ (as previously we
identify isomorphic and dual posets) and both posets are sortable in $C-c-1$
comparisons. Therefore we store the poset $P$ in the set ${\cal S}^*_c$ iff
both $P_1,P_2\in{\cal S}^*_{c+1}$ or $P_1\in{\cal S}^*_{c+1}$ and $P_2$ is
sortable in $C-c-1$ comparisons or $P_2\in{\cal S}^*_{c+1}$ and $P_1$ is
sortable in $C-c-1$ comparisons. Sortability of $P_1$ or $P_2$ can be checked
recursively using the same algorithm.

If some set ${\cal S}^*_c$ in the sequence appears to be empty then we conclude
that the poset $P_0$ cannot be sorted in $C$ comparisons. On the other hand, if
the set ${\cal S}^*_0$ is not empty, it contains the poset $P_0$ and we conclude
that the poset $P_0$ can be sorted in $C$ comparisons. For $n=13,14,15$ and
$C=C(n)$ we received that the set ${\cal S}^*_{15}$ is empty
\cite{Peczarski02,Peczarski04,Peczarski07}, which means that $S(n)>C(n)$ for
$n=13,14,15$. We analyze those results in detail in the next section.

\begin{figure}[p]
\begin{center}
\begin{tikzpicture}[scale=0.9]
\tikzstyle{vertex}=[circle, draw, minimum size=0.6cm, inner sep=0cm]
\foreach \n/\x/\y in {0/1.5/3, 1/0/2, 2/1/4, 3/2/0, 4/2/2,
                      5/3/1, 6/2/4, 7/2/1, 8/1/1, 9/1/0,
                      10/1/2, 11/0/1, 12/3/2}
  \node[vertex] (V-\n) at (\x, \y) {$u_{\n}$};
\foreach \from/\to in {0/2, 0/6, 1/0, 10/0, 4/0, 12/0,
                       11/10, 8/10, 7/4, 5/4, 9/8, 3/7}
  \draw (V-\from) -- (V-\to);
\end{tikzpicture}
\end{center}
\caption{The poset $P_{16}$, $e(P_{16})=113400$}
\label{fig:P16}
\end{figure}
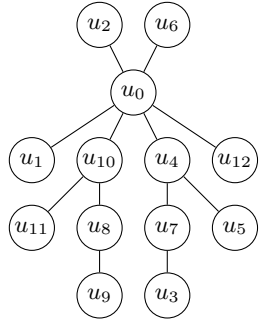

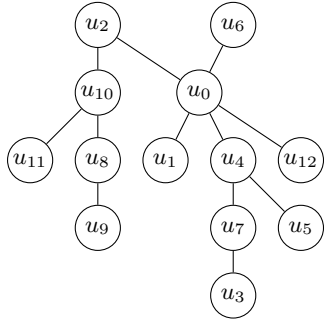
\begin{figure}[p]
\begin{center}
\begin{tikzpicture}[scale=0.9]
\tikzstyle{vertex}=[circle, draw, minimum size=0.6cm, inner sep=0cm]
\foreach \n/\x/\y in {0/2.5/3, 1/2/2, 2/1/4, 3/3/0, 4/3/2,
                      5/4/1, 6/3/4, 7/3/1, 8/1/2, 9/1/1,
                      10/1/3, 11/0/2, 12/4/2}
  \node[vertex] (V-\n) at (\x, \y) {$u_{\n}$};
\foreach \from/\to in {10/2, 0/2, 0/6, 1/0, 4/0, 12/0,
                       11/10, 8/10, 7/4, 5/4, 9/8, 3/7}
  \draw (V-\from) -- (V-\to);
\end{tikzpicture}
\end{center}
\caption{The poset $P'_{15}$, $e(P'_{15})=222750$}
\label{fig:P115}
\end{figure}

\begin{figure}[p]
\begin{center}
\begin{tikzpicture}[scale=0.9]
\tikzstyle{vertex}=[circle, draw, minimum size=0.6cm, inner sep=0cm]
\foreach \n/\x/\y in {0/2.5/3, 1/2/2, 2/1/5, 3/3/0, 4/3/2,
                      5/4/1, 6/3/4, 7/3/1, 8/1/3, 9/1/2,
                      10/1/4, 11/0/3, 12/4/2}
  \node[vertex] (V-\n) at (\x, \y) {$u_{\n}$};
\foreach \from/\to in {10/2, 0/10, 0/6, 1/0, 4/0, 12/0,
                       11/10, 8/10, 7/4, 5/4, 9/8, 3/7}
  \draw (V-\from) -- (V-\to);
\end{tikzpicture}
\end{center}
\caption{The poset $Q'_{16}$, $e(Q'_{16})=109350$}
\label{fig:Q116}
\end{figure}
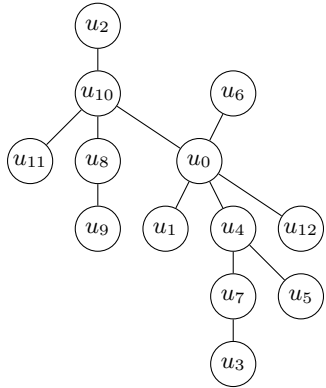

\section{The Previous Cases}\label{sec:prevcases}

\begin{figure}
\begin{center}
\begin{tikzpicture}[scale=0.9]
\tikzstyle{vertex}=[circle, draw, minimum size=0.6cm, inner sep=0cm]
\foreach \n/\x/\y in {0/1.5/4, 1/0/3, 2/1.5/5, 3/2/0, 4/2/2,
                      5/3/1, 6/2/3, 7/2/1, 8/1/2, 9/1/1,
                      10/1/3, 11/0/2, 12/3/3}
  \node[vertex] (V-\n) at (\x, \y) {$u_{\n}$};
\foreach \from/\to in {0/2, 6/0, 1/0, 10/0, 12/0, 4/6,
                       11/10, 8/10, 7/4, 5/4, 9/8, 3/7}
  \draw (V-\from) -- (V-\to);
\end{tikzpicture}
\end{center}
\caption{The poset $Q''_{16}$, $e(Q''_{16})=124740$}
\label{fig:Q216}
\end{figure}
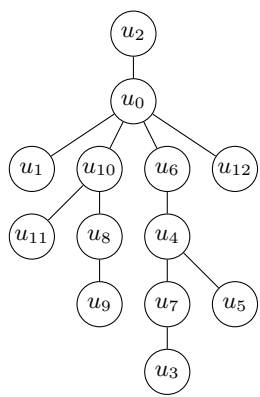

The computer experiment for $n=13$ and $C=C(n)$ returns that the set ${\cal
S}^*_{15}$ is empty, which means that $S(13)=F(13)=C(13)+1=34$
\cite{Peczarski02}. In that experiment the set ${\cal S}^*_{16}$ contains only
one poset $P_{16}$, whose Hasse diagram is shown in Fig.\ \ref{fig:P16}. The
poset $P_{16}$ can be obtained from a poset contained in the file ${\cal
S}_{15}$ in two ways:
\begin{itemize}
\item we compare elements $u_0$ and $u_{10}$ in the poset $P'_{15}\in{\cal
S}_{15}$ shown in Fig.\ \ref{fig:P115}; if $u_0>u_{10}$ we obtain the poset
$P_{16}$; if $u_0<u_{10}$ we obtain the poset $Q'_{16}$ shown in Fig.\
\ref{fig:Q116};
\item we compare elements $u_0$ and $u_{6}$ in the poset $P''_{15}\in{\cal
S}_{15}$ shown in Fig.\ \ref{fig:P215}; if $u_0<u_6$ we obtain the poset
$P_{16}$; if $u_0>u_6$ we obtain the poset $Q''_{16}$ shown in
Fig.\ \ref{fig:Q216}.
\end{itemize}
Neither the poset $P'_{15}$ nor the poset $P''_{15}$ can be stored in the file
${\cal S}^*_{15}$, because neither the poset $Q'_{16}$ nor the poset $Q''_{16}$
can be sorted in the remaining $C-16=17$ comparisons. It is quite surprising
that the posets $Q'_{16}$, $Q''_{16}$ cannot be sorted. The poset $Q'_{16}$ has
less linear extensions than the poset $P_{16}$, which intuitively should make it
easier to sort. Indeed, the poset $Q''_{16}$ has more linear extensions than the
poset $P_{16}$, which intuitively makes it harder to sort. On the other hand,
there are known the two largest elements of the poset $Q''_{16}$, which
intuitively makes it easier to sort. The poset $P_{16}$ is sortable in 17
comparisons because of its symmetry.

\begin{figure}
\begin{center}
\begin{tikzpicture}[scale=0.9]
\tikzstyle{vertex}=[circle, draw, minimum size=0.6cm, inner sep=0cm]
\foreach \n/\x/\y in {0/1.5/3, 1/0/2, 2/1.5/4, 3/3/0, 4/3/2,
                      5/4/1, 6/3/3, 7/3/1, 8/1/1, 9/1/0,
                      10/1/2, 11/0/1, 12/2/2}
  \node[vertex] (V-\n) at (\x, \y) {$u_{\n}$};
\foreach \from/\to in {0/2, 4/0, 1/0, 10/0, 12/0, 4/6,
                       11/10, 8/10, 7/4, 5/4, 9/8, 3/7}
  \draw (V-\from) -- (V-\to);
\end{tikzpicture}
\end{center}
\caption{The poset $P''_{15}$, $e(P''_{15})=238140$}
\label{fig:P215}
\end{figure}
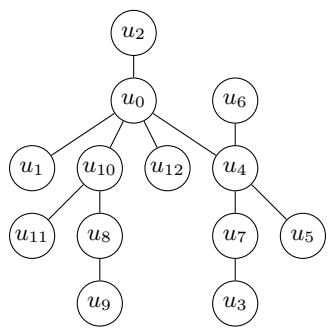

Similar results were received in the computer experiments for $n=14,15$ and
$C=C(n)$, i.e., $S(14)=F(14)=C(14)+1=38$ and $S(15)=F(15)=C(15)+1=42$. In both
cases the file ${\cal S}^*_{16}$ contains only one poset, namely the poset
$P_{16}$ extended by one isolated element $u_{13}$ (for $n=14$) or two isolated
elements $u_{13}$, $u_{14}$ (for $n=15$), respectively. In both cases the file
${\cal S}^*_{15}$ is empty and the reason is the same. The posets $P'_{15}$,
$Q''_{15}$, $Q'_{16}$, $Q''_{16}$ extended by $u_{13}$ or $u_{13}$, $u_{14}$ are
observed, respectively, and neither $Q'_{16}$ nor $Q''_{16}$ is sortable in the
remaining $C-16$ comparisons. Note that for $n=14$ we have $C-16=C(n)-16=21$ and
for $n=15$ we have $C-16=C(n)-16=25$.

\section{The Case of 16 Elements}\label{sec:attempt}

In this section we describe an attempt to find for $n=16$ a sorting algorithm
better than the FJA or to exclude existence of such algorithm. Before starting a
long time computation it was checked if the scenario from the previous section
repeats for $n=16$. The posets $Q'_{16}$, $Q''_{16}$ were extended by three
isolated elements $u_{13}$, $u_{14}$, $u_{15}$. As previously, the experiment
returned that neither the poset $Q'_{16}$ nor the poset $Q''_{16}$ can be sorted
in the remaining $C-16=C(n)-16=29$ comparisons. Of course, this result does not
exclude the existence of the desired algorithm.

To find the exact value of $S(16)$ the algorithm from Sect.\ \ref{sec:algorithm}
with improvement from Sect.\ \ref{sec:counting} is applied. Because the search
space is very reach, the problem is divided into smaller subproblems. Let $T(k)$
be the number of elements which were compared (touched) by a sorting algorithm
in the first $k$ comparisons. Observe that $T(k_1) \le T(k_2)$ for $k_1 < k_2$.
A sorting algorithm for 16 elements, using at most $C(16)=F(16)-1=45$
comparisons, is examined for possible values of $T(k)$.

The first experiment returned that if $S(16)=45$ then it holds $T(15)<16$. Note
that for the FJA we have $T(k)=16$ for $k\ge8$. Hence a hypothetical algorithm,
using for 16 elements pessimistically less comparisons than the FJA, must be
complete different from the FJA. It must differ from the FJA already before the
9th comparison. This is quite surprising, when we look at regular structure of
the first 15 comparisons in the FJA. The next experiment showed that if
$S(16)=45$ then $T(15)>11$, which is already not surprising.

\section{Computation Complexity}\label{sec:complexity}

Computation complexity of the method groves exponentially. The case $S(13)$
needed in year 2002 \cite{Peczarski02} more than 10 hours of CPU time. The value
of $S(14)$ was computed one year later (published in 2004 \cite{Peczarski02})
and took about 392 hours on faster computer and using improved algorithm, which
could solve $S(13)$ in about 40 minutes. Further progress in hardware allowed to
compute the value of $S(15)$ in year 2004 (published only in 2007
\cite{Peczarski07}) using about 17500 hours of CPU time. Each next case required
significant improvements in the algorithm or hardware. The progress is presented
in Table \ref{tab:times}. One can argue that the comparison is not fair, because
the machines used in the experiments are different. The purpose of this table is
to show an overall improvement in software and hardware, and to give a filling,
how difficult the case of 16 elements could be. The about 10 times improvement
observed between the second last and the last column is due mainly to the
algorithm described in the next section. Note that for Core 2 Due processor both
cores were used in parallel.

\begin{table}
\caption{Computation times}
\label{tab:times}
\begin{center}
\begin{tabular}{l*{4}{@{\quad}l}}
\hline
\noalign{\smallskip}
$n$ & Pentium II & Pentium III & Opteron 246 & Core 2 Duo \\
    & 233 MHz    & 650 MHz     & 2 GHz       & 2.13 GHz   \\
    & 2002       & 2003        & 2004        & 2007       \\
\noalign{\smallskip}
\hline
\noalign{\smallskip}
13 & 10 hr. 30 min. & 41 min.         & 10 min. 44 sec. & 46 sec.       \\
14 &                & 391 hr. 37 min. & 44 hr. 10 min.  & 4 hr. 31 min. \\
15 &                &                 & 17554 hr.       &               \\
\hline
\end{tabular}
\end{center}
\end{table}

A few years of CPU time was used up to now to search for an algorithm achieving
the ITLB for $n=16$. The computation that $T(15)<16$ and $T(15)>11$ took
about 20000 and 7000 hours, respectively. Computation for the next case
$T(15)=12$ is currently in progress. It used up to now more than 11000 hours.

\section{Counting Linear Extensions}\label{sec:counting}

The most time consuming part of the algorithm presented in Sect.\
\ref{sec:algorithm} is counting linear extensions of a given poset. In this
section we describe the algorithm for counting linear extensions which is
inspired by \cite{DeDeDe} and which substantially improves computations. For a
given poset $P=(U,\prec)$ the algorithm computes $e(P)$ and the table
$t[j,k]=e(P+u_j u_k)$ for $j \neq k$.

\begin{figure}
\begin{center}
\begin{tikzpicture}[scale=0.9]
\tikzstyle{vertex}=[circle, draw, minimum size=0.6cm, inner sep=0cm]
\tikzstyle{downset}=[circle, fill, inner sep=0.05cm]
\node[vertex] (V-a) at (-1,1.5) {$u_0$};
\node[vertex] (V-b) at (0, 1.5) {$u_1$};
\node[vertex] (V-c) at (-1, 2.5) {$u_2$};
\node[vertex] (V-d) at (0, 2.5) {$u_3$};
\draw (V-a) -- (V-c);
\draw (V-b) -- (V-d);
\draw (V-b) -- (V-c);
\node[downset] (D-0) at (5, 0) [label=below:{$\emptyset$}] {};
\node[downset] (D-2) at (4, 1) [label=left:{$\{u_1\}$}] {};
\node[downset] (D-1) at (6, 1) [label=right:{$\{u_0\}$}] {};
\node[downset] (D-a) at (3, 2) [label=left:{$\{u_1,u_3\}$}] {};
\node[downset] (D-3) at (5, 2) [label=right:{$\{u_0,u_1\}$}] {};
\node[downset] (D-b) at (4, 3) [label=left:{$\{u_0,u_1,u_3\}$}] {};
\node[downset] (D-7) at (6, 3) [label=right:{$\{u_0,u_1,u_2\}$}] {};
\node[downset] (D-f) at (5, 4) [label=above:{$\{u_0,u_1,u_2,u_3\}$}] {};
\foreach \from/\to in {0/2, 0/1, 2/a, 2/3, 1/3, a/b, 3/b, 3/7, b/f, 7/f}
  \draw[arrows=-latex] (D-\from) -- (D-\to);
\end{tikzpicture}
\end{center}
\caption{A poset and the graph of its downsets}
\label{fig:N}
\end{figure}
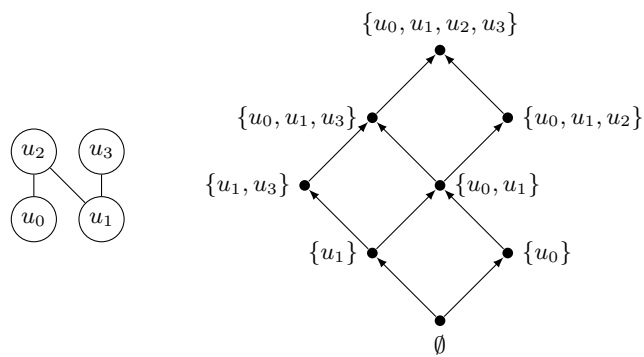

\begin{figure}
\begin{center}
\begin{tikzpicture}[scale=0.9]
\tikzstyle{downset}=[circle, fill, inner sep=0.05cm]
\node[downset] (D-0) at (5, 0) [label=below:{$d(\emptyset)=1,u(\emptyset)=5$}]
{};
\node[downset] (D-2) at (4, 1) [label=left:{$d(u_1)=1,u(u_1)=3$}] {};
\node[downset] (D-1) at (6, 1) [label=right:{$d(u_0)=1,u(u_0)=2$}] {};
\node[downset] (D-a) at (3, 2) [label=left:{$d(u_1,u_3)=1,u(u_1,u_3)=1$}] {};
\node[downset] (D-3) at (5, 2) [label=right:{$d(u_0,u_1)=2,u(u_0,u_1)=2$}] {};
\node[downset] (D-b) at (4, 3)
[label=left:{$d(u_0,u_1,u_3)=3,u(u_0,u_1,u_3)=1$}] {};
\node[downset] (D-7) at (6, 3)
[label=right:{$d(u_0,u_1,u_2)=2,u(u_0,u_1,u_2)=1$}] {};
\node[downset] (D-f) at (5, 4) [label=above:{$d(U)=5,u(U)=1$}] {};
\foreach \from/\to in {0/2, 0/1, 2/a, 2/3, 1/3, a/b, 3/b, 3/7, b/f, 7/f}
  \draw[arrows=-latex] (D-\from) -- (D-\to);
\end{tikzpicture}
\end{center}
\caption{The numbers of linear extensions of the downsets and they
complementary sets}
\label{fig:duN}
\end{figure}
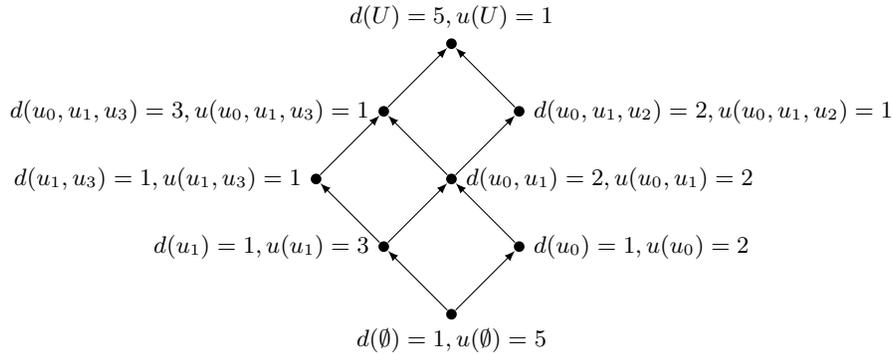

Let $P=(U,\prec)$ be a poset. A subset $D \subseteq U$ is called a down set of
the poset $P$ if for each $x \in D$ all elements $y \in U$ preceding $x$ (i.e.,
$y \prec x$) also belong to $D$. We consider a directed acyclic graph $\cal G$
whose nodes are all downsets of $P$. For two nodes $D_1$ and $D_2$ there is an
edge $(D_1,D_2)$ if there exists $x \in U \setminus D_1$ such that
$D_2=D_1\cup\{x\}$. An example of a poset and its graph of downsets is shown in
Fig.\ \ref{fig:N}, where $U=\{u_0,u_1,u_2,u_3\}$.

Let $d(D)$ denote the number of linear extensions of the poset $(D,\prec)$
which is the poset $P$ reduced to the down set $D$. Let $u(D)$ denote the
number of linear extensions of the poset $(U \setminus D,\prec)$ which is the
poset $P$ reduced to the complementary set of the down set $D$. We have
\cite{DeDeDe}
\begin{displaymath}
d(D) = \sum_{(X,D)}d(X),
\end{displaymath}
where the sum is taken over all edges $(X,D)$ in the graph $\cal G$ incoming to
the node $D$. We assume $d(\emptyset)=1$. Observe that $d(U)=e(P)$. All values
of $d(D)$ are computed using the DFS in the graph $\cal G$, starting at the node
$U$ and going down, i.e., in the opposite direction to the edges. Similarly, it
holds \cite{DeDeDe}
\begin{displaymath}
u(D) = \sum_{(D,X)}u(X),
\end{displaymath}
where the sum is taken over all edges $(D,X)$ in the graph $\cal G$ outgoing
from the node $D$. We assume $u(U)=1$. Observe that $u(\emptyset)=e(P)$. All
values of $u(D)$ are computed using the second DFS in the graph $\cal G$,
starting at the node $\emptyset$ and going up. Values of $d(D)$ and $u(D)$ for
the graph in Fig.\ \ref{fig:N} are shown in Fig.\ \ref{fig:duN}. The curly
braces are omitted for clarity, e.g., instead of $d(\{u_0\})$ we write $d(u_0)$.

\begin{table}
\caption{The values of $t[j,k]$}
\label{tbl:tN}
\begin{center}
\begin{tabular}{cc@{\quad}|*{4}{@{\quad}c}}
\hline
&&\multicolumn{4}{c}{$k$}\\
&&0&1&2&3\\
\hline
\multirow{4}{1em}{$j$}&0&--&2&5&4\\
&1&3&--&5&5\\
&2&0&0&--&2\\
&3&1&0&3&--\\
\hline
\end{tabular}
\end{center}
\end{table}

The table $t$ can be computed from the equation \begin{displaymath} t[j,k] =
\sum_{(V,W)}d(V)u(W), \end{displaymath} where the sum is taken over all edges
$(V,W)$ in the graph $\cal G$ such that $W=V\cup\{u_j\}$ and $u_k \in U
\setminus W$. For a proof see \cite{DeDeDe}. This computation is done altogether
with the second DFS. For the graph in Fig.\ \ref{fig:N} the values $t[j,k]$ are
included in Table \ref{tbl:tN}.

For a given poset on an $n$-element set its graph of downsets can have up to
$2^n$ nodes. We implemented the graph as a table of the size $2^n$. The table
is indexed by downsets. The index is the characteristic function of the set
$D$, i.e., the index is the $n$-bit number, where bit $j$ is set iff $u_j \in
D$. Graph $\cal G$ is not constructed explicitly. When we proceed a node $D$
all incoming and outgoing edges are easily computable from a poset
representation. We hold at position $D$ in the table only two numbers $d(D)$,
$u(D)$ and visited time stamp $v(D)$ needed to implement the DFS. We initialize
the table only once at the beginning of the program by setting all $v(D)=0$. We
also hold the global visited time stamp $v_t$ initialized to 0. Starting a new
DFS we increment the time stamp $v_t$. If we proceed a node $D$ and $v_t>v(D)$
then it means that the node $D$ was not yet visited in the current DFS run. If
$v_t=v(D)$ then the node was already visited. We do not need to reinitialize
the table before the next DFS. This is very important and decreases running
time. The algorithm is very efficient for small $n$, because with a high
probability the whole graph resides in a processor cache memory.

\end{document}